\documentclass[preprint,prd,superscriptaddress,showpacs,floatfix,%
preprintnumbers,nofootinbib]{revtex4}
\usepackage{color}
\usepackage{graphics}
\usepackage{amsmath, amssymb, epsfig}
\newcommand{\mathsym}[1]{{}}

\newcommand{\ba}{\begin{array}}
\newcommand{\ea}{\end{array}}
\newcommand{\be}{\begin{equation}}
\newcommand{\ee}{\end{equation}}
\newcommand{\bea}{\begin{eqnarray}}
\newcommand{\eea}{\end{eqnarray}}

\begin{document}
\vspace*{1cm}
\title{  Pseudo-Dirac neutrinos via mirror-world and depletion of UHE  neutrinos}
\bigskip
\author{Anjan S. Joshipura}
\email{anjan@prl.res.in}
\affiliation{Physical Research Laboratory, Navarangpura, Ahmedabad 380009,
India.}
\author{Subhendra Mohanty}
\email{mohanty@prl.res.in}
\affiliation{Physical Research Laboratory, Navarangpura, Ahmedabad 380009,
India.}
\author{Sandip Pakvasa}
\email{pakvasa@phys.hawaii.edu}
\affiliation{ Department of Physics \& Astronomy, University of Hawaii, Honolulu, HI 96822, USA\vspace*{1cm}}

%\preprint{TIFR/TH/13-03}
%\pacs{11.30.Hv, 14.60.Pq, 14.60.St}

\begin{abstract}
We propose a possible particle physics   explanation of the non-observation of muon neutrino events at IceCube coincident with GRB gamma ray at the rates  predicted by the standard Bahcall-Waxman model, in terms of neutrino oscillations. Our model is based on assuming that (a)  all neutrinos are pseudo-Dirac particles and (b) there exists a mirror world  interacting gravitationally with   the observed world. This scenario has three sterile neutrinos associated with each flavour of ordinary neutrinos. Very tiny
 mass splitting between these neutrinos is assumed to arise from lepton number violating dimension five operators suppressed by the Planck scale. We show that if  a mass splitting of $10^{-15} {\rm eV^2}$ is  induced between the four mass eigenstates of a given species, then its flux will be suppressed at IceCube energies by a factor of 4 which could be the explanation of the IceCube observation that the muon neutrino flux is lower than expected. Hierarchies in mass splitting among different flavours may result in different amount of suppression of each flavour and based on this, we predict a difference in the flavour ratios of the observed neutrinos which is significantly different compared to the standard prediction of $\nu_e:\nu_\mu:\nu_\tau=1:1:1$ which could serve as a test for this model.
\end{abstract}

\maketitle

\section{Introduction}

 Gamma Ray Burst (GRB) are believed \cite{Waxman:1995vg, Vietri:1995hs,  Milgrom:1995um} to be responsible for  accelerating  the charged cosmic rays  to very high energies and a sizable fluence of  neutrinos is expected from  GRB's  through the  interactions of  protons  with photons in the fireball \cite{Meszaros:1993tv}. The $p\gamma$ interaction  produce    charged pions:
$p+\gamma \rightarrow \Delta^+ \rightarrow n +\pi^+$ and subsequently the decays $\pi^+ \rightarrow \mu^+ \nu_\mu$ and
$\mu^+ \rightarrow  e^+ \nu_e \bar\nu_\mu$  produce muon and electron neutrinos coincident in direction with $\gamma$-rays.
The energy spectrum of these neutrinos are expected to peak in the range $10^5-10^7$GeV and in km$^3$ sized detectors like IceCube  and ANTARES about 10-100 neutrino events per year, coincident with GRB photons, were predicted \cite{Waxman:1997ti,Rachen,Waxman:1997ti2,Guetta,Ahlers}.  Contrary to expectations, no muon neutrino events coincident with GRB photons have been detected in measurements at IceCube \cite{Abbasi:2011qc, Abbasi:2012zw} which looks for the CC produced muons from $\nu_\mu$ through the Cerenkov radiation of upward going muons. An experimental upper bound on the muon neutrino flux from GRB's smaller than the Waxman-Bahcall prediction \cite{Waxman:1997ti,Guetta,Ahlers} by a factor 3.7 has been established with the IceCube observations \cite{Abbasi:2012zw}. Other experiments
probing the ultra-high-energy regime, such as ANITA\cite{vieregg}
have not seen any evidence of PeV energy neutrinos in association with GRB events either. A recent observation \cite{Fan:2013zoa} of a GRB at a low redshift of $z\sim 0.34$ and an estimated isotropic electromagnetic energy of $10^{54}$ makes it the most energetic of  GRB's emission seen at $z<0.5$. No neutrinos coincident with this GRB event within 100 $s$ and $3.5^{\circ}$ were seen at IceCube \cite{IceCube130427A}. A search for muon neutrinos associated with GRB's performed with the ANTARES detector \cite{Adrian-Martinez:2013sga} shows no events over the atmospheric neutrino background.

 There are  two PeV neutrino events which have been seen at IceCube \cite{Aartsen:2013bka}. They  are expected to be of cosmogenic origin \cite{Barger:2012mz}. There are a further 26 events seen at IceCube at energies between 0.02-0.3 PeV \cite{Whitehorn}. These 28 events may be initiated by neutrinos from dark matter decay \cite{Feldstein:2013kka, Barger:2013pla} or the neutrinos may have a astrophysical origin \cite{Laha}. There is only one muon track in the 10 events between energies 0.1-1.15 PeV \cite{Barger:2013pla} and the paucity of muon neutrinos continues to be evident  even in these non-GRB related events suggesting possible non-astrophysical origin for the muon neutrino depletion.

  In the light of the IceCube non-observation of the GRB neutrinos the standard GRB fireball or internal shock model parameters have been revised
  \cite{Hummer:2011ms,LiZh,He:2012tq,Zhang:2012qy,Gao:2013fra} in a full numerical calculation  by taking into account in greater detail dilution of charged pions and kaons in the expanding fireball and due to multi-pion and kaon production. The new neutrino flux limits are consistent with the upper bound put from
 IceCube  \cite{Abbasi:2011qc, Abbasi:2012zw, IceCube130427A}.
  Other ways of explaining the paucity of IceCube neutrinos without overthrowing the GRB models is to explain it by oscillations of the flavor neutrinos into sterile ones.
There exist strong constraints on the possible oscillations of active to sterile neutrinos from the terrestrial experiments \cite{Abazajian:2012ys, Giunti:2011gz, Giusarma:2011ex}  and from
nucleosynthesis \cite{Shi:1993hm,Enqvist:1991qj,Kirilova:1999xj}. But due to very high energy and long distances, the relevant mass scale probed through the active sterile oscillations of UHE are completely different
and are as yet unconstrained. For example, consider the neutrinos from GRB's (at cosmological distances $L=1000-3000$Mpc) and  in the range $E=10^5 - 10^7$GeV.
These active UHE neutrinos ($\nu_\alpha, \alpha=e, \mu, \tau$)  can significantly oscillate to sterile neutrinos  ($\nu^\prime$) if the  (mass)$^2$ difference between them is
\be\label{grbdletam}
\Delta m^2_{\nu_\alpha \nu^\prime} \geq 0.8 \times 10^{-15} {\rm eV^2} \left(\frac{1000 {\rm Mpc}}{L}\right)\left(\frac{E}{10^7 {\rm GeV}} \right)
\ee
and mixing between them is significant. A pair of active and a sterile neutrino with the above (mass)$^2$ difference may be regarded as a pseudo-Dirac neutrino which in some limit can be considered  a   Dirac particle  respecting some unbroken Lepton number. Small violation of this symmetry not only splits them but also mix them maximally.
UHE neutrinos can thus be used to probe the pseudo-Dirac nature of neutrinos \cite{Kobayashi:2000md, Beacom:2003eu,deGouvea:2009fp,Esmaili:2009fk,Barry:2010en, Esmaili:2012ac,pakvasa}. The maximal mixing within a pseudo Dirac pair can bring a suppression in the original flux by a factor of 1/2. Note that this is over and above the suppression in muon neutrino flux resulting from the active to active oscillation which is already included in the  the Waxman-Bahcall prediction \cite{Waxman:1997ti}  of the neutrino flux.  This falls significantly short of the suppression
by 3.7  observed
at IceCube. If $\Delta m^2_{\nu_\alpha \nu^\prime}$ is in the range $10^{-18}-10^{-16} {\rm eV}^2$ the interference terms in the oscillation probability does not average to zero for the range of red-shifts of the observed GRBs and the suppression factor can be lower than 1/2 \cite{Esmaili:2012ac}. Other explanations offered for the IceCube muon neutrino deficit are neutrino spin flip in a magnetic field \cite{Barranco:2012xj} and neutrino decays over cosmological distances \cite{pakvasa,Baerwald:2012kc}.

 We discuss here how a  suppression by a factor of $\sim$ 1/4 can be achieved in the flux of UHE  (a) if  all neutrinos are pseudo-Dirac and (b) if there exists a  mirror world \cite{Foot:1995pa,Berezhiani:1995yi,Berezinsky:2002fa} interacting gravitationally with   the observed world. Global lepton number breaking through the gravitational interactions provides a source in  this scenario which can split the mass eigenstates of all the Dirac neutrinos and make them pseudo-Dirac.
\section{Pseudo-Dirac neutrinos via mirror world}
First we consider the  case of a single flavour say, $\nu_\mu$. Assume that  $\nu_\mu$ is a Dirac particle
and is accompanied by a mirror ``muon neutrino'' also a Dirac particle. They together consist of
four two component left handed states labeled as $\nu^\prime_{\mu a}$ with $a=1...4$. Of these, $\nu^\prime_{\mu 1}$ is  active and others are sterile states. Their mass matrix  to the leading order is given by
\begin{equation}
{\mathcal M}^0_\mu=m_{\nu_\mu}\left( \begin{array}{cccc}
0&1&0&0\\
1&0&0&0\\
0&0&0&1\\
0&0&1&0\\
\end{array}\right)
\end{equation}
We have assumed the same mass for the both the Dirac states.
All the zeros in the above mass matrix are protected by Lepton numbers in our and the mirror worlds. Breaking of these symmetries is assumed to induce  small entries in  places of zeros. Thus for example, non-zero values for 11 and its mirror symmetric 33 elements are induced by the conventional dimension five Weinberg operator in our and mirror world respectively  and may arise from the seesaw mechanism or gravity induced effects.
Following the mirror world scenario \cite{Foot:1995pa}, we assume that
ordinary and mirror worlds communicate only gravitationally with each other, thus $(\nu^\prime_{\mu 1},\nu^\prime_{\mu3})$ get coupled only gravitationally through the dimension-5 operator
\cite{Foot:1995pa,Berezhiani:1995yi, Berezinsky:2002fa}:
\begin{equation}
{\cal L}_{comm}= \frac{\lambda_{ 13}}{M_P} (\nu^\prime_{\mu 1} \phi) (\nu^\prime_{\mu 3}\phi^\prime)
\end{equation}
where $\phi$ and $\phi^\prime$ are the neutral components of the Higgses in our and the mirror world respectively.
The contribution of this dimension 5 operator to the mass matrix is
\be \label{strength}
\epsilon_{13}\equiv\frac{\lambda_{ 13} v^2}{M_P}
\ee
where we have taken   $v=\langle \phi \rangle\simeq\langle \phi^\prime \rangle\simeq 174$GeV.
The sterile partners $(\nu^\prime_{\mu 2}$ and $\nu^\prime_{\mu 4}$) may couple to different set of Higgs $\eta$ and $\eta^\prime$ and assuming that these Higgs vevs are at the TeV scale, one can have a gravitational mixing term from
all the sterile pairs. For example
$(\nu^\prime_{ \mu 2},\nu^\prime_{\mu 4})$ will mix via the operator
\begin{equation}
{\cal L^\prime}_{comm}= \frac{\lambda _{24}}{M_P} (\nu^\prime_{\mu 2} \eta) (\nu^\prime_{\mu 4}\eta^\prime)
\end{equation}
where $\eta$ and $\eta^\prime$ are the neutral components of the Higgses in our and the mirror world respectively.
The contribution of this dimension 5 operator to the mass matrix is
\be \label{strength2}
\epsilon_{24}=\frac{\lambda_{24}  }{M_P}\langle \eta \rangle \langle \eta^\prime \rangle
\ee
Taking similar terms for mixing of all sterile pairs, we can write the mass matrix in the $\nu^\prime_{\mu a}$ basis
as
\begin{equation} \label{M2}
{\mathcal M_\mu}=\left(
\begin{array}{cc cc}
 \epsilon_{11}&m_{\nu_\mu}&\epsilon_{13}&\epsilon_{14}\\
m_{\nu_\mu}&\epsilon_{22}&\epsilon_{23}&\epsilon_{24}\\
\epsilon_{13}&\epsilon_{23}&\epsilon_{11}&m_{\nu_\mu}\\
\epsilon_{14}&\epsilon_{24}&m_{\nu_\mu} &\epsilon_{22}\\
\end{array}
\right)
\end{equation}
We have  assumed that the ordinary and mirror worlds are symmetric if their mixing is neglected and thus assumed $\epsilon_{33}=\epsilon_{11}$ and $\epsilon_{22}=\epsilon_{44}$ in eq.( \ref{M2}). We also assume that  $\epsilon$ parameters are $ \ll m_{\nu_\mu}$ and neutrinos remain pseudo-Dirac.
Let
\begin{equation}
{\mathcal V}^T_4 {\mathcal M_\mu}{\mathcal  V_4}\equiv {\rm diag}(m_{\mu1},m_{\mu2},m_{\mu3},m_{\mu4})~, \end{equation}
Eigenvalues $m_{\mu a}$ are given to leading orders in $\epsilon$ as
\begin{eqnarray}
\label{ev}
 m_{\mu1}&\simeq& \frac{1}{2} (2 m_{\nu_\mu}+\epsilon_{11}+\epsilon_{13}+\epsilon_{14}+\epsilon_{22}+\epsilon_{23}+
 \epsilon_{24})~, \nonumber \\
 m_{\mu2}&\simeq& \frac{1}{2} (-2 m_{\nu_\mu}+\epsilon_{11}+\epsilon_{13}-\epsilon_{14}+\epsilon_{22}-\epsilon_{23}+
 \epsilon_{24})~,\nonumber \\
 m_{\mu3}&\simeq& \frac{1}{2} (2 m_{\nu_\mu}+\epsilon_{11}-\epsilon_{13}-\epsilon_{14}+\epsilon_{22}-\epsilon_{23}-
 \epsilon_{24})~,\nonumber \\
 m_{\mu4}&\simeq& \frac{1}{2} (-2 m_{\nu_\mu}+\epsilon_{11}-\epsilon_{13}+\epsilon_{14}+\epsilon_{22}+\epsilon_{23}-
 \epsilon_{24})~.
\end{eqnarray}

Diagonalizing matrix is given to the same order by
\begin{equation}
\label{V}
 {\mathcal V_4}\equiv {\mathcal V}^0_4\overline{\mathcal V_4}\simeq \frac{1}{2}\left(
\begin{array}{cccc}
 1&-1&-1&1\\
1&1&-1&-1\\
1&-1&1&-1\\
1&1&1&1\\
\end{array}
\right)~\left(
\begin{array}{cccc}
 1&y_{1}&0&y_{2}\\
-y_{1}&1&y_{2}&0\\
0&-y_{2}&1&y_{3}\\
-y_{2}&0&-y_{3}&1\\
\end{array}
\right)~,
\end{equation}
where
\begin{eqnarray} \label{ys}
y_{1} &\equiv& \frac{1}{4 m_{\nu_\mu}}(\epsilon_{11}+\epsilon_{13}-\epsilon_{22}-\epsilon_{24}),\nonumber\\
y_{2} &\equiv& \frac{1}{4m_{\nu_\mu}}(\epsilon_{23}-\epsilon_{14}),\nonumber\\
y_{3} &\equiv& \frac{1}{4m_{\nu_\mu}}(\epsilon_{11}-\epsilon_{13}-\epsilon_{22}+\epsilon_{24}).
\end{eqnarray}
The role of $\epsilon_{ab}$ is essentially to split all four degenerate states and mixing between them is essentially determined by
the  $\epsilon_{ab}$ independent matrix ${\mathcal V}_4^0$ in eq.(\ref{V}). In the following, we shall assume that all the parameters
$\epsilon_{ab}$ have the same typical magnitude given by
\be \label{magnitude}
\epsilon_{ab}\equiv \epsilon\sim \frac{\lambda}{M_P} v^2\approx 2.4 \times 10^{-6} \lambda~~{\rm eV}~,\ee
where $v\sim 174$GeV. Then it follows from eq.(\ref{ev}) that a typical scale responsible for the long wavelength oscillations of muon
neutrinos is given by
\be \label{magnitude2}
\Delta_{2}\approx 2 m_{\nu_\mu} \epsilon\approx 4.5 \times 10^{-8}~ {\rm eV}^2 \lambda\left(\frac{m_{\nu_\mu}}{0.009~ {\rm eV}}\right)~.\ee

The oscillation length associated with this scale and  energy $E=10^5-10^7~{\rm GeV}$ is smaller than  the typical distance of the UHE sources,see eq.(\ref{grbdletam}) and the effect of $\Delta_2$ gets averaged out resulting in suppression of the muon neutrino flux.  The averaged survival probability is essentially determined by ${\mathcal V}_4^0$ in eq.(\ref{V}) and is given by:
\begin{equation}\label{p2}
P_{\mu \mu}=4 \left(\frac{1}{2}\right)^4=\frac{1}{4}
\end{equation}
This reduction is over and above the flux reduction which takes place due to averaged oscillations between active flavours and one needs to
generalize the above formulation to take this effect into account. We do this in the next section.

\section{Three generations}
In the following, we shall assume a straightforward generalization of the above scenario and require that all three active neutrinos and their mirror partners are pseudo-Dirac.  We are thus dealing with  12 left handed states
$\nu_{\alpha a}^\prime ,~ \alpha=e,\mu,\tau~, a=1..4$ in this case and mixing among them would now be governed by a $12\times 12$ matrix.
The mass matrix (\ref{M2}) for a single flavour is generalised to a $12\times 12$ matrix for the three flavours as follows. The mass $m_{\nu_\mu}$ is replaced by a $3\times3$ mass matrix $m^{\alpha \beta}$ in the flavour space $\alpha, \beta=e,\mu, \tau$,
\be
m_{\nu_\mu} \rightarrow m^{\alpha \beta}\equiv {\mathbf  m}
\ee
This matrix can be diagonalised by the usual bi-unitary transformation:
\be
 {\bf U_L}^T  {\mathbf  m}{\bf  U_R}={\rm  diagonal}~ (m_1,m_2,m_3)\equiv {\bf d_\nu}.
\ee
Here, the matrix ${\bf U_L}$ governing the left handed mixing can be identified with the usual MNSP matrix $ \equiv {\bf U}$.
Each of the the parameters $\epsilon_{ab}$ appearing in eq.(\ref{V}) now gets replaced by a $3\times 3$ matrices in flavour space
\be
 \epsilon_{ab}\rightarrow \epsilon^{\alpha \beta}_{ab}\equiv  {\bf \epsilon}_{ab}
\ee
These matrices are generated by dim. 5 operators  as before, .e.g. $\epsilon_{13}^{\alpha\beta}$ arise from  gravitational mixing between neutrinos in our universe and the mirror universe:
\begin{equation}
\frac{\lambda^{\alpha \beta}_{13}}{M_P} (\nu^\prime_{\alpha 1} \phi) (\nu^\prime_{\beta 3}\phi^\prime)
\end{equation}
which gives the $3 \times 3$ mixing matrix in flavour space,
\be \label{strength3}
\epsilon_{13}^{\alpha\beta}=\frac{\lambda^{\alpha \beta}_{13} v^2}{M_P}\approx
 2.5\times 10^{-6} \lambda^{\alpha \beta}_{ 13}\, {\rm eV}~.
\ee
%Similarly
%$(\nu^\prime_{ \alpha 2},\nu^\prime_{\beta 4})$ will mix via the operator
%\begin{equation}
%{\cal L^\prime_{\rm 3}}_{comm}= \frac{\lambda^{\alpha \beta}_{ 2 4}}{M_P} (\nu^\prime_{\alpha 2} \eta) (\nu^\prime_{\beta 4}\eta^\prime)
%\end{equation}
%and the contribution of this dimension 5 operator to the mass matrix can be written as the $3 \times 3$ matrix,
%\be \label{strength4}
%\epsilon^{\alpha \beta}_{2 4 }=\frac{\lambda^{\alpha \beta}_{ 2 4}  }{M_P}\langle \eta \rangle \langle \eta^\prime \rangle
%\ee
Taking similar terms for mixing of all  pairs, we write the $12 \times 12$ mass matrix in the $(\nu^\prime_{\alpha, 1},\nu^\prime_{\alpha, 2},\nu^\prime_{\alpha, 3},\nu^\prime_{\alpha, 4})$  basis as
\begin{equation} \label{M12}
{\mathcal M}=\left(
\begin{array}{cc cc}
 {\mathbf \epsilon}_{11}&{\mathbf m}&{\mathbf \epsilon}_{13}&{\mathbf \epsilon}_{14}\\
{\mathbf m}& {\mathbf \epsilon}_{22}&{\mathbf \epsilon}_{23}&{\mathbf \epsilon}_{24}\\
{\mathbf \epsilon}_{13}^T&{\mathbf \epsilon}_{23}^T& {\mathbf \epsilon}_{11}&{\mathbf m}\\
{\mathbf \epsilon}_{ 14}^T&{\mathbf \epsilon}_{24}^T&{\mathbf  m} &{\mathbf  \epsilon}_{22}\\
\end{array}
\right)
\end{equation}
Note that each entry above is a $3\times 3$ matrix in the generation space. This matrix can be diagonalised by the following steps. We first diagonalise the ${\mathbf  m}$ blocks by
the matrix,
\be \label{U12}
{\mathcal U}^\prime= \left(
\begin{array}{cc cc}
{\bf U_L}&{\bf 0} &{\bf 0}&{\bf 0}\\
{\bf  0}&{\bf U_R}&{\bf 0}&{\bf 0}\\
{\bf 0}&{\bf 0}&{\bf U_L}&{\bf 0}\\
{\bf 0}&{\bf 0}&{\bf 0}&{\bf U_R} \\
\end{array}
\right)
\end{equation}
with the transformation,
\begin{equation}\label{Mprime}
{\mathcal M}^\prime={\mathcal U^\prime}^T{\mathcal M U^\prime}=\left(
\begin{array}{cc cc}
 \tilde{{\mathbf \epsilon}}_{11}&{\bf d}_{\nu}&\tilde{{\mathbf \epsilon}}_{13}&\tilde{{\mathbf \epsilon}}_{14}\\
{\bf d}_{\nu}&\tilde{{\mathbf \epsilon}}_{22}&\tilde{{\mathbf \epsilon}}_{23}&\tilde{{\mathbf \epsilon}}_{24}\\
\tilde{{\mathbf \epsilon}}_{13}^T&\tilde{{\mathbf \epsilon}}_{23}^T&\tilde{{\mathbf \epsilon}}_{11}&{\mathbf d}_{\nu}\\
\tilde{{\mathbf \epsilon}}_{14}^T&\tilde{{\mathbf \epsilon}}_{24}^T&{\bf d}_{\nu} &\tilde{{\mathbf \epsilon}}_{22}\\
\end{array}
\right)
\end{equation}
where
\bea
\tilde{{\mathbf \epsilon}}_{ab}&\equiv& U_L^T {\mathbf \epsilon_{ab}}U_L~~~{\rm for}~~~~ ab=11,13~,\nonumber \\
\tilde{{\mathbf \epsilon}}_{ab}&\equiv& U_L^T {\mathbf \epsilon_{ab}}U_R~~~{\rm for}~~~~ ab=14~,\nonumber \\
\tilde{{\mathbf \epsilon}}_{ab}&\equiv& U_R^T {\mathbf \epsilon_{ab}}U_L~~~{\rm for}~~~~ ab=23~,\nonumber \\
\tilde{{\mathbf \epsilon}}_{ab}&\equiv& U_R^T {\mathbf \epsilon_{ab}}U_R~~~{\rm for}~~~~ ab=22,24~.
\eea
The diagonal elements   $(\tilde{{\mathbf \epsilon}}_{ab})^{ii}$ in flavour space  serve to split the masses $m_{\nu_i}$ of the $i^{th}$ flavour. The off diagonal entries give corrections to them
and also mix sterile states of different flavours. Since mixing  of  an active neutrino of one flavour with a sterile neutrino associated with a different flavour
is strongly constrained from experiments, we shall assume that off diagonal entries  of each of the matrices $\tilde{{\mathbf \epsilon}}_{ab}$ are small compared to the diagonal ones and take these matrices as diagonal:
\be
\tilde{\mathbf \epsilon}_{ab}^D\equiv  {\rm diagonal}(\epsilon^1_{ab},\epsilon^2_{ab} ,\epsilon^3_{ab} )~,
\ee
where $\tilde{\mathbf \epsilon}_{ab}^D$ are now diagonal $3\times 3$ matrix for each $ab$.
 In this approximation, eq. (\ref{Mprime}) can be written as
\begin{equation} \label{Mprime2}
{\mathcal M}^\prime\approx \left(
\begin{array}{cc cc}
\tilde{\mathbf \epsilon}_{11}^D&{\bf d_{\nu}}&\tilde{\mathbf  \epsilon}_{13}^D&\tilde{\mathbf  \epsilon} _{14}^D\\
{\bf d_{\nu}}&\tilde{\mathbf \epsilon}_{22}^D&\tilde{\mathbf\epsilon}_{23}^D&\tilde{ \mathbf \epsilon}_{24}\\
\tilde{\mathbf \epsilon}_{13}^D&\tilde{\mathbf \epsilon}_{23}^D&\tilde{\mathbf \epsilon}_{11}^D&{\bf d_{\nu}}\\
\tilde{\mathbf \epsilon}_{14}^D&\tilde{\mathbf  \epsilon}_{24}^D&{\bf d_{\nu}} &\tilde{\mathbf \epsilon}_{22}^D\\
\end{array}
\right)~,
\end{equation}
To the first order in $\tilde{\mathbf \epsilon}_{ab}^D{\bf d_\nu^{-1}}$, the matrix ${\mathcal M}^\prime$ is now  diagonalised by,
\begin{equation}
\label{Vprime}
 {\mathcal V}\equiv {\mathcal V}^0\overline{ \mathcal V}\simeq \frac{1}{2}\left(
\begin{array}{cccc}
 {\bf I}&-{\bf I}&-{\bf I}&{\bf I}\\
{\bf I}&{\bf I}&-{\bf I}&-{\bf I}\\
{\bf I}&-{\bf I}&{\bf I}&-{\bf I}\\
{\bf I}&{\bf I}&{\bf I}&{\bf I}\\
\end{array}
\right)~\left(
\begin{array}{cccc}
 {\bf I}&{\bf y_{1}}&{\bf 0}&{\bf y_{2}}\\
-{\bf y_{1}}& {\bf I}&{\bf y_{2}}&{\bf 0}\\
{\bf 0}&-{\bf y_{2}}&{\bf I}&{\bf y_{3}}\\
-{\bf y_{2}}&{\bf 0}&-{\bf y_{3}}&{\bf I}\\
\end{array}
\right)~,
\end{equation}
where ${\bf I}$ (${\bf 0}$)denotes $3\times 3$ identity (null)  matrix. The diagonal $3\times 3$ matrices ${\bf y_{1,2,3}}$ are given by expressions analogous to eq.(\ref{ys}):
\begin{eqnarray} \label{ys2}
{\bf y}_{1} &\equiv& \frac{1}{4}(\tilde{\mathbf\epsilon}_{11}^D+\tilde{\mathbf \epsilon}_{13}^D-\tilde{\mathbf  \epsilon}_{22}^D-\tilde{ \mathbf \epsilon}_{24}^D) {\bf d_\nu^{-1}},\nonumber\\
{\bf y}_{2} &\equiv& \frac{1}{4}(\tilde{\mathbf \epsilon}_{23}^D-\tilde{\mathbf \epsilon}_{14}^D){\bf d_\nu^{-1}},\nonumber\\
{\bf y}_{3} &\equiv& \frac{1}{4}(\tilde{\mathbf\epsilon}_{11}^D-\tilde{\mathbf \epsilon}_{13}^D-\tilde{\mathbf  \epsilon}_{22}^D+\tilde{ \mathbf \epsilon}_{24}^D) {\bf d_\nu^{-1}}~. \end{eqnarray}
%\bea
%{\bf y_{1}} \equiv \frac{1}{4} {\bf m_\nu^{-1}}({\bf \epsilon_{12}}-{\bf \epsilon_{24}}),\nonumber\\
%{\bf y_{2}} \equiv \frac{1}{4} {\bf m_\nu^{-1}} ({\bf \epsilon_{14}}-{\bf \epsilon_{23}}).\nonumber\\
%\eea
The 12 eigenvalues of ${\mathcal M}$ are given to leading order in $\tilde{\bf \epsilon}_{ab}$ by equation analogous to (\ref{ev}):
\begin{eqnarray}
\label{ev3}
 {\bf m}_{1}&\simeq& \frac{1}{2} (2 {\bf d}_{\nu}+\tilde{\mathbf \epsilon}_{11}^D+\tilde{\mathbf \epsilon}_{13}^D+\tilde{\mathbf \epsilon}_{14}^D+\tilde{\mathbf \epsilon}_{22}^D+\tilde{\mathbf \epsilon}_{23}^D+\tilde{\mathbf \epsilon}_{24}^D)
\nonumber \\
{\bf m}_{2}&\simeq& \frac{1}{2} (-2 {\bf d}_{\nu}+\tilde{\mathbf \epsilon}_{11}^D+\tilde{\mathbf \epsilon}_{13}^D-\tilde{\mathbf \epsilon}_{14}^D+\tilde{\mathbf \epsilon}_{22}^D-\tilde{\mathbf \epsilon}_{23}^D+\tilde{\mathbf \epsilon}_{24}^D)
\nonumber \\
{\bf m}_{3}&\simeq& \frac{1}{2} (2 {\bf d}_{\nu}+\tilde{\mathbf \epsilon}_{11}^D-\tilde{\mathbf \epsilon}_{13}^D-\tilde{\mathbf \epsilon}_{14}^D+\tilde{\mathbf \epsilon}_{22}^D-\tilde{\mathbf \epsilon}_{23}^D-\tilde{\mathbf \epsilon}_{24}^D)
\nonumber \\
{\bf m}_{4}&\simeq& \frac{1}{2} (-2 {\bf d}_{\nu}+\tilde{\mathbf \epsilon}_{11}^D-\tilde{\mathbf \epsilon}_{13}^D+\tilde{\mathbf \epsilon}_{14}^D+\tilde{\mathbf \epsilon}_{22}^D+\tilde{\mathbf \epsilon}_{23}^D-\tilde{\mathbf \epsilon}_{24}^D)
~.\end{eqnarray}

The mixing matrix ${\mathcal U}$ between the 12 gauge eigenstates $(\nu_{\alpha 1}^\prime,\nu_{\alpha 2}^\prime, \nu_{\alpha 3}^\prime,\nu_{\alpha 4}^\prime),~(\alpha=e, \mu,\tau$) and the mass eigenstates $(\nu_{i 1},\nu_{i 2}, \nu_{i 3},\nu_{i 4}), i=1\cdots 3$ is given by the product of ${\mathcal U}'$  eq.(\ref{U12}) and ${\mathcal V}$,eq.(\ref{Vprime}).
To zeroeth order in ${\bf \epsilon_{ab}}$, one can approximate ${\mathcal V}$ by ${\mathcal V}^0$   and ${\mathcal U}$ is approximately given
by
\begin{equation}
\label{Vfin}
{\mathcal U}\equiv {\mathcal U}^\prime {\mathcal V}\simeq \frac{1}{2}\left(
\begin{array}{cccc}
 {\bf U_L}&-{\bf U_L}&-{\bf U_L}&{\bf U_L}\\
{\bf U_R}&{\bf U_R}&-{\bf U_R}&-{\bf U_R}\\
{\bf U_L}&-{\bf U_L}&{\bf U_L}&-{\bf U_L}\\
{\bf U_R}&{\bf U_R}&{\bf U_R}&{\bf U_R}\\
\end{array}
\right)
\end{equation}
In this approximation, the three flavour eigenstates $\nu_\alpha\equiv \nu_{\alpha 1}^\prime$ are given in terms of 12 mass mass eignestates
$\nu_{ia}$ from the above equation by
\be \label{flavour}
\nu_\alpha\equiv \nu_{\alpha 1}^\prime={\mathcal U}^{\alpha i}_{1 a}\, \nu_{i a}\equiv \frac{1}{2} {\bf U}^{\alpha i} (\nu_{i1}-\nu_{i 2}-\nu_{i 3}+\nu_{i4})
\ee
with ${\bf U_L}\equiv {\bf U}$ denoting the MNSP matrix. The mass $m_{ia}$ of each component $\nu_{ia}$ is given  by ${\bf (m_a)}_{ii}$ from eq.(\ref{ev3}). The splitting among the mirror partners of a given mass eigenstate $\nu_{i,a}$  is then given by
$\Delta_{ab}^{i}\equiv m_{ia}^2-m_{ib}^2$.
The corresponding oscillation probabilities follow from the time evolution of the state $\nu_i$ defined in eq. (\ref{flavour}):
 \bea
 \label{prob}
 P_{\alpha\beta}(L)&=& \frac{1}{16} \sum_{ij} {\bf U}^*_{\alpha i}{\bf U}_{\beta i}{\bf U}_{\beta j}{\bf U}^*_{\alpha j} e^{-i(m^2_{j 1}-m^2_{i 1})\frac{L}{2E}}\nonumber\\
&\times& \left(1+ e^{i\chi^i_{12}}+e^{i\chi^i_{13}}+e^{i\chi^i_{14}}\right)\left(1+ e^{-i\chi^j_{12}}+e^{-i\chi^j_{13}}+e^{-i\chi^j_{14}}\right)
 \eea
 where $\chi^i_{ab}\equiv \Delta^{i}_{ab}\frac{L}{2 E}.$
The four states $\nu_{ia}$ for a given $i$ are degenerate when $\Delta^{i}_{ab}$ are small and $\chi^i_{ab}$ can be neglected as in typical short baseline experiments.
 In this limit, $\nu_i$ defined in eq.(\ref{flavour}) behave as a single mass eigenstate and one recovers the standard mixing and oscillations of the flavour states.   The $\Delta^i_{ab}$ induce observable   long wavelength oscillations between active and sterile states. For long baselines with ($E/L> 10^{-12} {\rm eV^2}$) the exponential factor
$e^{-i(m^2_{j 1}-m^2_{i 1})\frac{L}{2E}}$  in (\ref{prob})  averages to zero if $i \neq j$. The oscillation probability in  the  long baseline experiments then can be written as,
\bea
 \label{prob2}
 P_{\alpha\beta}(L)= \frac{1}{8} \sum_{i} {|\bf U}_{\alpha i}|^2{|\bf U}_{\beta i}|^2\,
 (2&+&  \cos\chi^i_{12}+\cos\chi^i_{13}+\cos\chi^i_{14} \nonumber\\
 &+&\cos\chi^i_{23}+\cos\chi^i_{24}+\cos\chi^i_{34})
 \eea

Typical scales associated with these splittings can be written assuming normal mass hierarchy as
\begin{eqnarray} \label{dmall}
\Delta^1_{ab}&\simeq&  4 \lambda_1 m_1   \frac{v^2}{M_p^2}\ll  9 \times 10^{-8}\,\lambda_1\,\, {\rm eV^2}~,\nonumber \\
\Delta^2_{ab}& \simeq & 4 \lambda_2 \sqrt{\Delta_\odot}  \frac{v^2}{M_p^2}=  9 \times 10^{-8}\,\lambda_2\,\, {\rm eV^2}~,\nonumber \\
\Delta^3_{ab} &\simeq&  4 \lambda_3 \sqrt{\Delta_{atm}}  \frac{v^2}{M_p^2}=  4.7 \times 10^{-7}\,\lambda_3\,\, {\rm eV^2}~,\end{eqnarray}
where $\lambda_{1,2,3}$ denote the gravitational couplings in three sectors  controlling the strength of the dim. 5 operators.
 These  are  constrained from two major considerations. The number of  neutrino species in equilibrium at the time of  BBN  is severely constrained. Requirement that a  sterile neutrino does  not equilibrate at that time through large angle oscillations to active one implies that  their  (mass)$^2$ difference must obey  \cite{Shi:1993hm,Enqvist:1991qj,Kirilova:1999xj}
$\Delta m^2_{\nu_\alpha \nu^\prime} \leq 10^{-9} {\rm eV}^2$.
A stronger constrain exist on $\Delta^1_{ab}$. In the approximation of neglecting mixing between active and sterile partners of different generations, $\Delta^1_{ab}$ control the solar neutrino oscillations. The corresponding oscillation length for MeV neutrino is of the order of the Sun-Earth distance for $\Delta^1_{ab}\sim 10^{-12}$ eV$^2$. Such $\Delta^1_{ab}$ can modify the LMA solution and detailed fits in case of pseudo-Dirac neutrinos \cite{Berezinsky:2002fa}
imply a bound $\Delta^1_{ab}<1.8\times  10^{-12}$ eV $^2$ at 3$\sigma$.  One expects similar but somewhat stronger bound when mirror partners are also present. This  bound can be satisfied either by choosing  $m_1\ll \sqrt{\Delta_\odot}$ or in case of
$m_1 \sim O(\sqrt{\Delta_\odot})$ by choosing $\lambda_1\leq 10^{-5}$.

 We will assume that all the splittings among a given flavour $\Delta^{ii}_{ab}$ for different pairs of $ab$ are equal
  and in this case the oscillation probability (\ref{prob2}) reduces to the simple form
   \bea
 \label{prob3}
 P_{\alpha\beta}(L)= \frac{1}{4} \sum_{i} {|\bf U}_{\alpha i}|^2{|\bf U}_{\beta i}|^2\,
 (1+3 \cos\chi_i)
 \eea
 where $\chi_i\equiv \chi^i_{ab}\, \forall a,b$. One can now work out the observed flux ratios of UHE neutrinos using  this $P_{\alpha\beta}$. The flux $\Phi_\beta=(\phi_e,\phi_\mu,\phi_\tau)$ in a flavour $\beta$ is given by
\be
\label{flux}
\Phi_\beta=P_{\beta\alpha}\Phi^0_\alpha \ee
where $\Phi^0_\alpha $ denotes the initial flux. For $\Phi^0_\alpha\sim\frac{\phi_0}{3} (1,2,0)$ one obtains
\be \label{flux2}
\Phi_\beta\sim \frac{\phi_0}{12} \sum_i|{\bf U}_{\beta i}|^2(|{\bf U}_{e i}|^2+2|{\bf U}_{\mu i}|^2)(1+3 \cos\chi_i)~.
\ee
One recovers the standard value $\Phi_\beta^{S}=\frac{\phi_0}{3} \sum_i|{\bf U}_{\beta i}|^2(|{\bf U}_{e i}|^2+2|{\bf U}_{\mu i}|^2)$ with only three Dirac neutrinos in the limit $\chi_i=0$. The deviation in flux compared to the standard value is thus given by
\be
\label{difference}
\delta\Phi_\beta\equiv\Phi_\beta-\Phi_\beta^S=-\frac{\phi_0}{2}\sum_i|{\bf U}_{\beta i}|^2(|{\bf U}_{e i}|^2+2|{\bf U}_{\mu i}|^2)\sin^2\frac{\chi_i}{2}~.\ee
For maximal atmospheric mixing and $\theta_{13} \simeq 0$, $|{\bf U}_{e i}|^2+2|{\bf U}_{\mu i}|^2=1$ for every $i$  and the above simplifies to
\be
\label{differecne2}
\delta\Phi_\beta=-\frac{\phi_0}{2}\sum_i|{\bf U}_{\beta i}|^2\sin^2\frac{\chi_i}{2}~.\ee
This is to be compared with the corresponding formula\cite{Beacom:2003eu, pakvasa}  obtained for the pseudo-Dirac  neutrinos in the absence of
the mirror neutrinos:
\be
\delta\Phi_\beta=-\frac{\phi_0}{3}\sum_i|{\bf U}_{\beta i}|^2\sin^2\frac{\chi_i}{2}~.
\ee
The presence of pseudo-Dirac mirror partners now lead to stronger deviation in $ \Phi_\beta$  from the canonical value 1/3. The values of
$\delta \Phi_\beta$ depend both on the values of $|{\bf U}_{\beta i}|$ which are now reasonably well-known and on the hierarchies in $\Delta_i$ given typically by.  eq.(\ref{dmall}).
There exist two interesting ranges of $\Delta_i$ which can effect the oscillations of UHE in  physically different ways:\\
\noindent (A) Strong mass hierarchies among neutrinos $m_1\ll m_2\simeq \sqrt{\Delta_{odot}}<m_3\simeq \sqrt{\Delta_{atm}}$ ( or equivalently
$ \lambda_1\ll \lambda_2\simeq \lambda_{2,3}$ in eq.(\ref{dmall})) such that $\Delta_{1}$ in eq.(\ref{dmall}) is $<10^{-16}$ eV$^2$  but $\Delta_{2,3}>10^{-16}$ eV$^2$. The $\Delta_1$ in this case  does not induce the appreciable oscillations of the UHE neutrinos while effects of  $\Delta_{2,3}$ can be averaged out.  This corresponds to taking $\chi_1=\cos\chi_2=\cos\chi_3=0$ in eq.(\ref{flux2}) and one obtains
$$\phi_\beta\approx \frac{\phi_0}{12}(1+3 |U_{\beta 1}|^2)$$
which  translates to
\be
\label{fluxratio1}
\phi_e:\phi_\mu:\phi_\tau\approx 2:1:1 ~\ee
for the tri-bimaximal mixing. The corresponding number  for the current best fit values \cite{global} of mixing angles  is $2.12:1:1.09$.
While fluxes in all three flavours are suppressed compared to the canonical value 1/3, the suppression of the electron neutrinos flux is less.\\
\noindent (B) The alternative possibility corresponds to a milder hierarchy characterized by $m_1\approx \sqrt{\Delta_{\odot}}$ and all $\lambda_i$
similar in magnitude such that $\Delta_1$ is suppressed compared to $\Delta_{2,3}$ to satisfy the solar bound  but all of them  still are bigger than
 the oscillation scale $\sim 10^{-16}$ eV$^2$ of the UHE neutrinos. This case corresponds to taking $\cos\chi_i=0$ for all $i$ in eq.(\ref{flux2}) and all the flavours are suppressed by a factor of 4 compared to the canonical value of 1/3.

Although no neutrinos of any flavour have been seen from GRB's, the fact that  of the nine IceCube events in the  0.15 -1.15 PeV range there is only one with a muon track suggests a preferential suppression of muon neutrinos, which indicates the hierarchial mass splitting of scenario (A).

In our model we have introduced 9 extra neutrinos which can potentially be in conflict with the BBN constraints on the effective number of species of light particles during nucleosynthesis. Of these extra neutrinos, $\nu_2^\alpha, \nu_4^\alpha$ ($\alpha=e,\mu,\tau$) are sterile and can decouple much before the time of BBN, their temperatures will be smaller than the radiation bath and they will not contribute to the Hubble expansion at the time of BBN. In our model  intergenerational mixing between an active neutrino of one flavour with sterile neutrinos associated with other is negligible. As a result,   $\nu_2,\nu_3$ will not equilibriate with the active $\nu_1,\nu_3$ species by oscillation. There is no equilibrium attained by $\nu_1 \leftrightarrow \nu_{2,3,4}$ oscillations of the same flavour as the mass splittings  is $\le 10^{-9} {\rm eV^2}$ \cite{Shi:1993hm,Enqvist:1991qj,Kirilova:1999xj}.  However $\nu_3^\alpha$ are 'active' in the mirror world and they could count as three extra neutrino degrees of freedom if their temperature were to be identical to the temperature of the active neutrinos in our world. One way to avoid this doubling of neutrino degrees is to assume that the mirror world couplings to the inflaton are slightly different and the reheating temperature of the mirror world following inflation is  lower than reheat temperature of our universe \cite{Berezhiani:1995am}.
The effective neutrino degrees of freedom observed by Planck \cite{planck} at the time of matter-radiation equality is
$N_{eff}=3.30 \pm 0.27$ at 68\% CL. If there are $N_{m}$ species of mirror neutrinos with temperature $T_m$ then they will count as
\be
N_{eff}=3.046 + N_{m} \left(\frac{T_{m}}{T_\nu}\right)^4
\ee
neutrinos. We see that in order that $\Delta N_{eff} < 0.3$ with  $N_m=9$ extra mirror neutrino species the mirror neutrino
 temperature $T_m <0.43 T_\nu$ at the time of matter-radiation equality to evade the Planck bound. Another way in which sterile neutrinos, including mirror ones, can evade the stringent Planck bound is if there is an annihilation of
 MeV dark matter preferentially into photons such that the photon temperature relative to the neutrino temperature
is raised after neutrino decoupling and prior to $z_{eq}$. Scenarios for evading the Planck constraint on sterile neutrinos via dark matter models have been discussed in ref. \cite{scherrer, Steigman}

\section{Conclusions}
 In this paper we have studied the neutrino fluxes and flavour ratios of $10^5-10^7$GeV neutrinos originating from the GRB in a scenario
 with three sterile neutrinos for each flavour having   tiny splitting as  given in eq.(\ref{grbdletam}) among them.
It is shown that  in this scenario  GRB neutrinos of all flavours  or muon and tau flavour can get suppressed
by factor of 1/4 as required by IeCube result.  This suppression can result in  the presence of maximal  mixing among a neutrino and three
sterile partners as given in eq.(\ref{V}). Such mixing  can arise if
all the  neutrinos are pseudo-Dirac and there is a mirror world replicating our own interactions.
As far as $\Delta_e$ is concerned, it is required to be  $ < 10^{-12} {\rm eV^2}$ \cite{Berezinsky:2002fa}. This  leads to an interesting possibility.
UHE neutrinos are also expected  at energies of the
GZK limit cosmic rays with energies of $10^9$ GeV and their sources are closer at distances of 100 Mpc. Thus the (mass)$^2$ difference required for significant conversion of these neutrinos  should be $\geq 10^{-12}$ eV$^2$. Thus it is possible that electron neutrinos from GRB get depleted but one from the nearby sources and high energy remain undepleted. Similar thing   can happen for other flavours also if $\lambda_\mu,\lambda_\tau$ in eq.(\ref{dmall}) are such that  $\Delta_{\mu,\tau}$ also
fall in the range  $10^{-12}-10^{-15}$ eV$^2$. These could serve as discriminating tests of models of pseudo-Dirac neutrinos like the one discussed in this paper.\\ \\
\noindent{\bf Acknowledgements}\\ \\
ASJ thanks the Department of Science and Technology, Government of India for support under the J. C.
Bose National Fellowship programme, grant no. SR/S2/JCB-31/2010 and  SP  and SM acknowledge
the  support of  the Indo-US Joint Centre on
{\it Physics Beyond the Standard Model}-JC/23-2010, and from the U.S.D.O.E. under the grant DE-FC02-04ER41291. SP would like to thank the Alexander von Humboldt Foundation for support
and Professot Heinrich Paes and the department of Physics at the
University
of Dortmund for hospitality while this work was being completed.


\begin{thebibliography}{99}
\bibitem{Waxman:1995vg}
  E.~Waxman,
  %``Cosmological gamma-ray bursts and the highest energy cosmic rays,''
  Phys.\ Rev.\ Lett.\  {\bf 75}, 386 (1995)
  [astro-ph/9505082].
  %%CITATION = ASTRO-PH/9505082;%%
\bibitem{Vietri:1995hs}
  M.~Vietri,
  %``On the acceleration of ultrahigh-energy cosmic rays in gamma-ray bursts,''
  Astrophys.\ J.\  {\bf 453}, 883 (1995)
  [astro-ph/9506081].
  %%CITATION = ASTRO-PH/9506081;%%
%\cite{Milgrom:1995um}
\bibitem{Milgrom:1995um}
  M.~Milgrom and V.~Usov,
  %``Possible association of ultrahigh-energy cosmic ray events with strong gamma-ray bursts,''
  Astrophys.\ J.\  {\bf 449}, L37 (1995)
  [astro-ph/9505009].
  %%CITATION = ASTRO-PH/9505009;%%

\bibitem{Meszaros:1993tv}
  P.~Meszaros and M.~J.~Rees,
  %``Relativistic fireballs and their impact on external matter - Models for cosmological gamma-ray bursts,''
  Astrophys.\ J.\  {\bf 405}, 278 (1993).
  %%CITATION = ASJOA,405,278;%%

\bibitem{Waxman:1997ti}
  E.~Waxman and J.~N.~Bahcall,
  %``High-energy neutrinos from cosmological gamma-ray burst fireballs,''
  Phys.\ Rev.\ Lett.\  {\bf 78}, 2292 (1997)
  [astro-ph/9701231].
\bibitem{Rachen}  J.~P.~Rachen and P.~Meszaros,
  %``Cosmic rays and neutrinos from gamma-ray bursts,''
  AIP Conf.\ Proc.\  {\bf 428}, 776 (1997)  [astro-ph/9811266].  %%CITATION = ASTRO-PH/9811266;%%
\bibitem{Waxman:1997ti2}E.~Waxman and J.~N.~Bahcall,
  %``High-energy neutrinos from astrophysical sources: An Upper bound,''
  Phys.\ Rev.\ D {\bf 59}, 023002 (1999)
  [hep-ph/9807282].
\bibitem{Guetta}D. Guetta et al., Astropart Phys. 20,
429 (2004).

\bibitem{Ahlers}M. Ahlers, M.C. Gonzalez-Garcia and F.
Halzen, Astroparticle Phys. 35, 87 (2011).

\bibitem{Abbasi:2011qc}
  R.~Abbasi {\it et al.}  [IceCube Collaboration],
  %``Limits on Neutrino Emission from Gamma-Ray Bursts with the 40 String IceCube Detector,''
  Phys.\ Rev.\ Lett.\  {\bf 106}, 141101 (2011)
  [arXiv:1101.1448 [astro-ph.HE]].
  %%CITATION = ARXIV:1101.1448;%%

\bibitem{Abbasi:2012zw}
  R.~Abbasi {\it et al.}  [IceCube Collaboration],
  %``An absence of neutrinos associated with cosmic-ray acceleration in $\gamma$-ray bursts,''
  Nature {\bf 484}, 351 (2012)
  [arXiv:1204.4219 [astro-ph.HE]].
  %%CITATION = ARXIV:1204.4219;%%
 %\cite{Abbasi:2011qc}


\bibitem{vieregg} A. G. Vieregg et al., Astrophys. J. 736, 50 (2011);
  arXiv:1102.3206.

%\bibitem{auger}  Pierre Auger Collaboration (P. Abreu et al.),
 % Astrophys. J. 755, L4 (2012); arXiv:1210.3143.



\bibitem{Fan:2013zoa}
  Y.~-Z.~Fan, P.~H.~T.~Tam, F.~-W.~Zhang, Y.~-F.~Liang, H.~-N.~He, B.~Zhou, R.~-Z.~Yang and Z.~-P.~Jin {\it et al.},
  %``High energy emission of GRB 130427A: evidence for inverse Compton radiation,''
  arXiv:1305.1261 [astro-ph.HE].
  %%CITATION = ARXIV:1305.1261;%%
  %3 citations counted in INSPIRE as of 16 Jun 2013

\bibitem{IceCube130427A}http://gcn.gsfc.nasa.gov/gcn3/14520.gcn3


%\bibitem{Ishihara} A. Ishihara, �Ice Cube: Ultra High Energy Neutrinos�,
%talk at Neutrino 2012, Kyoto Japan, June 2012; slides
%available at http://neu2012.kek.jp/index.html.

%\cite{Adrian-Martinez:2013sga}
\bibitem{Adrian-Martinez:2013sga}
  S.~Adri�n-Mart�nez, A.~Albert, I.~A.~Samarai, M.~Andr�, M.~Anghinolfi, G.~Anton, S.~Anvar and M.~Ardid {\it et al.},
  %``Search for muon neutrinos from gamma-ray bursts with the ANTARES neutrino telescope using 2008 to 2011 data,''
  arXiv:1307.0304 [astro-ph.HE].
  %%CITATION = ARXIV:1307.0304;%%


\bibitem{Aartsen:2013bka}
  M.~G.~Aartsen {\it et al.}  [IceCube Collaboration],
  %``First observation of PeV-energy neutrinos with IceCube,''
  arXiv:1304.5356 [astro-ph.HE].
  %%CITATION = ARXIV:1304.5356;%%
  %9 citations counted in INSPIRE as of 15 Jun 2013

\bibitem{Barger:2012mz}
  V.~Barger, J.~Learned and S.~Pakvasa,
  %``IceCube PeV Cascade Events Initiated by Electron-Antineutrinos at Glashow Resonance,''
    arXiv:1207.4571 [astro-ph.HE].  %%CITATION = ARXIV:1207.4571;%%  %5 citations counted in INSPIRE as of 19 Feb 2013
%\cite{Barger:2013pla}

\bibitem{Whitehorn}N. Whitehorn, C. Kopper, and N. Neilson. (for IceCube Collaboration), Talk at IceCube Particle Astrophysics Symposium, at Madiosn, Wisconsin, USA, May 15, 2013.

\bibitem{Feldstein:2013kka}
  B.~Feldstein, A.~Kusenko, S.~Matsumoto and T.~T.~Yanagida,
  %``Neutrinos at IceCube from Heavy Decaying Dark Matter,''
  arXiv:1303.7320 [hep-ph].
  %%CITATION = ARXIV:1303.7320;%%
  %2 citations counted in INSPIRE as of 15 Jun 2013

\bibitem{Barger:2013pla}
  V.~Barger and W.~-Y.~Keung,
  %``Superheavy Particle Origin of IceCube PeV Neutrino Events,''
  arXiv:1305.6907 [hep-ph].
  %%CITATION = ARXIV:1305.6907;%%
  %1 citations counted in INSPIRE as of 15 Jun 2013

\bibitem{Laha} R.~Laha, J.~FBeacom, B.~Dasgupta, S.~Horiuchi and K.~Murase,
  %``Demystifying the PeV Cascades in IceCube: Less (Energy) is More (Events),''
  arXiv:1306.2309 [astro-ph.HE].
  %%CITATION = ARXIV:1306.2309;%%

\bibitem{Hummer:2011ms}
  S.~Hummer, P.~Baerwald and W.~Winter,
  %``Neutrino Emission from Gamma-Ray Burst Fireballs, Revised,''
  Phys.\ Rev.\ Lett.\  {\bf 108}, 231101 (2012)  [arXiv:1112.1076 [astro-ph.HE]].  %%CITATION = ARXIV:1112.1076;%%  %20 citations counted in INSPIRE as of 13 Feb 2013

\bibitem{LiZh}
Z.~Li,
  %``Note on the Normalization of Predicted GRB Neutrino Flux,''
   [arXiv:1112.2240 [astro-ph.HE]].  %%CITATION = ARXIV:1112.2240;%%  %15 citations counted in INSPIRE as of 18 Mar 2013
  Z.~Li, Phys. Rev. D85,02730(2012),
  %``Fermi Limit of the Neutrino Flux from Gamma-ray Bursts,''
   arXiv:1210.6594 [astro-ph.HE].  %%CITATION = ARXIV:1210.6594;%%  %1 citations counted in INSPIRE as of 19 Feb 2013



\bibitem{He:2012tq}
  H.~-N.~He, R.~-Y.~Liu, X.~-Y.~Wang, S.~Nagataki, K.~Murase and Z.~-G.~Dai,
  %``Icecube non-detection of GRBs: Constraints on the fireball properties,''
    Astrophys.\ J.\  {\bf 752}, 29 (2012)  [arXiv:1204.0857 [astro-ph.HE]].  %%CITATION = ARXIV:1204.0857;%%  %15 citations counted in INSPIRE as of 19 Feb 2013

\bibitem{Zhang:2012qy}
  B.~Zhang and P.~Kumar,
  %``Model-dependent high-energy neutrino flux from Gamma-Ray Bursts,''
  Physical Review Letters, vol.\  110, Issue 12, id.\  {\bf 121101} (2013)
  [arXiv:1210.0647 [astro-ph.HE]].
  %%CITATION = ARXIV:1210.0647;%%
  %6 citations counted in INSPIRE as of 15 Jun 2013

%\bibitem{Baerwald:2013pu}
%  P.~Baerwald, M.~Bustamante and W.~Winter,
  %``UHECR escape mechanisms for protons and neutrons from GRBs, and the cosmic ray-neutrino connection,''
 % arXiv:1301.6163 [astro-ph.HE].  %%CITATION = ARXIV:1301.6163;%%

%\bibitem{Gao:2012ay}
 % S.~Gao, K.~Asano and P.~Meszaros,
  %``High Energy Neutrinos from Dissipative Photospheric Models of Gamma Ray Bursts,''
  % JCAP {\bf 1211}, 058 (2012)  [arXiv:1210.1186 [astro-ph.HE]].  %%CITATION = ARXIV:1210.1186;%%  %2 citations counted %in INSPIRE as of 13 Feb 2013

%\bibitem{Murase:2013hh}
 % K.~Murase, K.~Kashiyama and P.~Meszaros,
  %``Subphotospheric Neutrinos from Gamma-Ray Bursts: The Role of Neutrons,''
  %  arXiv:1301.4236 [astro-ph.HE].  %%CITATION = ARXIV:1301.4236;%%

%\cite{Gao:2013fra}
\bibitem{Gao:2013fra}
  S.~Gao, K.~Kashiyama and P.~Meszaros,
  %``On the neutrino non-detection of GRB 130427A,''
  arXiv:1305.6055 [astro-ph.HE].
  %%CITATION = ARXIV:1305.6055;%%
  %1 citations counted in INSPIRE as of 16 Jun 2013

\bibitem{Abazajian:2012ys}
  K.~N.~Abazajian {\it et al.},
  %, M.~A.~Acero, S.~K.~Agarwalla, A.~A.~Aguilar-Arevalo, C.~H.~Albright, S.~Antusch, C.~A.~Arguelles and A.~B.~Balantekin ,
  %``Light Sterile Neutrinos: A White Paper,''
   arXiv:1204.5379 [hep-ph].
     %%CITATION = ARXIV:1204.5379;%%
\bibitem{Giunti:2011gz}
  C.~Giunti and M.~Laveder,
  %``3+1 and 3+2 Sterile Neutrino Fits,''
    Phys.\ Rev.\ D {\bf 84}, 073008 (2011)  [arXiv:1107.1452 [hep-ph]].  %%CITATION = ARXIV:1107.1452;%%
\bibitem{Giusarma:2011ex}
  E.~Giusarma, M.~Corsi, M.~Archidiacono, R.~de Putter, A.~Melchiorri, O.~Mena and S.~Pandolfi,
  %``Constraints on massive sterile neutrino species from current and future cosmological data,''
   Phys.\ Rev.\ D {\bf 83}, 115023 (2011)  [arXiv:1102.4774 [astro-ph.CO]].  %%CITATION = ARXIV:1102.4774;%%

\bibitem{Shi:1993hm}
  X.~Shi, D.~N.~Schramm and B.~D.~Fields,
  %``Constraints on neutrino oscillations from big bang nucleosynthesis,''
  Phys.\ Rev.\ D {\bf 48}, 2563 (1993)
  [astro-ph/9307027].
  %%CITATION = ASTRO-PH/9307027;%%
\bibitem{Enqvist:1991qj}
  K.~Enqvist, K.~Kainulainen and M.~J.~Thomson,
  %``Stringent cosmological bounds on inert neutrino mixing,''
  Nucl.\ Phys.\ B {\bf 373}, 498 (1992).
  %%CITATION = NUPHA,B373,498;%%
\bibitem{Kirilova:1999xj}
  D.~P.~Kirilova and M.~V.~Chizhov,
  %``Cosmological nucleosynthesis and active sterile neutrino oscillations with small mass differences: The Resonant case,''
   Nucl.\ Phys.\ B {\bf 591}, 457 (2000)  [hep-ph/9909408].  %%CITATION = HEP-PH/9909408;%%


\bibitem{Kobayashi:2000md}
  M.~Kobayashi and C.~S.~Lim,
  %``Pseudo Dirac scenario for neutrino oscillations,''
  Phys.\ Rev.\ D {\bf 64}, 013003 (2001)
  [hep-ph/0012266].
 \bibitem{Beacom:2003eu}
 J.~F.~Beacom, N.~F.~Bell, D.~Hooper, J.~G.~Learned, S.~Pakvasa and T.~J.~Weiler,
  %``PseudoDirac neutrinos: A Challenge for neutrino telescopes,''
  Phys.\ Rev.\ Lett.\  {\bf 92}, 011101 (2004)
  [hep-ph/0307151].



 %\cite{deGouvea:2009fp}
\bibitem{deGouvea:2009fp}
  A.~de Gouvea, W.~-C.~Huang and J.~Jenkins,
 % ``Pseudo-Dirac Neutrinos in the New Standard Model,''
 Phys.\ Rev.\ D {\bf 80}, 073007 (2009)  [arXiv:0906.1611 [hep-ph]].  %%CITATION = ARXIV:0906.1611;%%


\bibitem{Esmaili:2009fk}
  A.~Esmaili,
  %``Pseudo-Dirac Neutrino Scenario: Cosmic Neutrinos at Neutrino Telescopes,''
  Phys.\ Rev.\ D {\bf 81}, 013006 (2010)
  [arXiv:0909.5410 [hep-ph]].
  %%CITATION = ARXIV:0909.5410;%%
\bibitem{Barry:2010en}
  J.~Barry, R.~N.~Mohapatra and W.~Rodejohann,
  %``Testing the Bimodal/Schizophrenic Neutrino Hypothesis in Neutrino-less Double Beta Decay and Neutrino Telescopes,''
  Phys.\ Rev.\ D {\bf 83}, 113012 (2011)
  [arXiv:1012.1761 [hep-ph]].
  %%CITATION = ARXIV:1012.1761;%%

 \bibitem{Esmaili:2012ac}
  A.~Esmaili and Y.~Farzan,
  %``Implications of the Pseudo-Dirac Scenario for Ultra High Energy Neutrinos from GRBs,''
   JCAP {\bf 1212}, 014 (2012)  [arXiv:1208.6012 [hep-ph]].  %%CITATION = ARXIV:1208.6012;%%

\bibitem{pakvasa}
  S.~Pakvasa, A.~Joshipura and S.~Mohanty, Phys.\ Rev.\ Lett.\  {\bf 110}, 171802(2013).
  %``Explanation for the low flux of high energy astrophysical muon-neutrinos,''
  arXiv:1209.5630 [hep-ph].

\bibitem{Barranco:2012xj}
  J.~Barranco, O.~G.~Miranda, C.~A.~Moura and A.~Parada,
  %``A reduction in the UHE neutrino flux due to neutrino spin precession,''
  Phys.\ Lett.\ B {\bf 718}, 26 (2012)  [arXiv:1205.4285 [astro-ph.HE]].  %%CITATION = ARXIV:1205.4285;%%

\bibitem{Baerwald:2012kc}
  P.~Baerwald, M.~Bustamante and W.~Winter,
  %``Neutrino Decays over Cosmological Distances and the Implications for Neutrino Telescopes,''
    JCAP {\bf 1210}, 020 (2012)  [arXiv:1208.4600 [astro-ph.CO]].  %%CITATION = ARXIV:1208.4600;%%








\bibitem{Foot:1995pa}
  R.~Foot and R.~R.~Volkas,
  %``Neutrino physics and the mirror world: How exact parity symmetry explains the solar neutrino deficit, the atmospheric neutrino anomaly and the LSND experiment,''
  Phys.\ Rev.\ D {\bf 52}, 6595 (1995)
  [hep-ph/9505359].
  %%CITATION = HEP-PH/9505359;%%

 \bibitem{Berezhiani:1995yi}
  Z.~G.~Berezhiani and R.~N.~Mohapatra,
  %``Reconciling present neutrino puzzles: Sterile neutrinos as mirror neutrinos,''
   Phys.\ Rev.\ D {\bf 52}, 6607 (1995)  [hep-ph/9505385].  %%CITATION = HEP-PH/9505385;%%

 \bibitem{Berezinsky:2002fa}
  V.~Berezinsky, M.~Narayan and F.~Vissani,
  %``Mirror model for sterile neutrinos,''
  Nucl.\ Phys.\ B {\bf 658}, 254 (2003)
  [hep-ph/0210204].
  %%CITATION = HEP-PH/0210204;%%


 \bibitem{global}
 %\cite{Tortola:2012te}
%\bibitem{Tortola:2012te}
  D.~V.~Forero, M.~Tortola and J.~W.~F.~Valle,
  %``Global status of neutrino oscillation parameters after Neutrino-2012,''
  Phys.\ Rev.\ D {\bf 86}, 073012 (2012)
  [arXiv:1205.4018 [hep-ph]];
  %%CITATION = ARXIV:1205.4018;%%
%\cite{Fogli:2012ua}
%\bibitem{Fogli:2012ua}
  G.~L.~Fogli, E.~Lisi, A.~Marrone, D.~Montanino, A.~Palazzo and A.~M.~Rotunno,
  %``Global analysis of neutrino masses, mixings and phases: entering the era of leptonic CP
%violation searches,''
  Phys.\ Rev.\ D {\bf 86}, 013012 (2012)
  [arXiv:1205.5254 [hep-ph]];
  %%CITATION = ARXIV:1205.5254;%%;
  M.~C.~Gonzalez-Garcia, M.~Maltoni, J.~Salvado and T.~Schwetz,
  %``Global fit to three neutrino mixing: critical look at present precision,''
  JHEP {\bf 1212}, 123 (2012)
  [arXiv:1209.3023 [hep-ph]].
  %%CITATION = ARXIV:1209.3023;%%
\bibitem{Berezhiani:1995am}
  Z.~G.~Berezhiani, A.~D.~Dolgov and R.~N.~Mohapatra,
  %``Asymmetric inflationary reheating and the nature of mirror universe,''
  Phys.\ Lett.\ B {\bf 375}, 26 (1996)  [hep-ph/9511221].

\bibitem{planck}
  P.~A.~R.~Ade {\it et al.}  [ Planck Collaboration],
  %``Planck 2013 results. XVI. Cosmological parameters,''
  arXiv:1303.5076 [astro-ph.CO].  %%CITATION = ARXIV:1303.5076;%%  %9 citations counted in INSPIRE as of 28 Mar 2013



\bibitem{scherrer}
  C.~M.~Ho, R.~J.~Scherrer and ,
  %``Limits on MeV Dark Matter from the Effective Number of Neutrinos,''
  Phys.\ Rev.\ D {\bf 87}, 023505 (2013)  [arXiv:1208.4347 [astro-ph.CO]];  %%CITATION = ARXIV:1208.4347;%%
  C.~M.~Ho, R.~J.~Scherrer ,
  %``Sterile Neutrinos and Light Dark Matter Save Each Other,''
   Phys.\  Rev.\  D 87, {\bf 065016} (2013)  [arXiv:1212.1689 [hep-ph]].

%\cite{Steigman:2013yua}
\bibitem{Steigman}
  G.~Steigman,
  %``Equivalent Neutrinos, Light WIMPs, and the Chimera of Dark Radiation,''
  Phys.\ Rev.\ D {\bf 87}, 103517 (2013)
  [arXiv:1303.0049 [astro-ph.CO]].
  %%CITATION = ARXIV:1303.0049;%%
  %7 citations counted in INSPIRE as of 16 Jun 2013





  %\cite{Donini:2008xn}
%\bibitem{Donini:2008xn}
 % A.~Donini and O.~Yasuda,
  %``Signatures of sterile neutrino mixing in high-energy cosmic neutrino flux,''
 % arXiv:0806.3029 [hep-ph].
  %%%CITATION = ARXIV:0806.3029;%%






\end{thebibliography}
\end{document}